\documentclass[11pt,preprint]{aastex}
\slugcomment{Accepted to PASP}

\def\kms{\ifmmode{\rm km\thinspace s^{-1}}\else km\thinspace s$^{-1}$\fi}
\def\ms{\ifmmode{\rm m\thinspace s^{-1}}\else m\thinspace s$^{-1}$\fi}

\shortauthors{S. Howell}
\shorttitle{Fringe Science}

\begin{document}

\title{Fringe Science: \\
Defringing CCD Images with Neon Lamp Flat Fields
}

\author{
Steve~B.~Howell\altaffilmark{1} \\
NASA Ames Research Center, Moffett Field, CA 94035
}

\altaffiltext{1}{Visiting Astronomer, Kitt Peak National Observatory, 
National Optical Astronomy Observatory, which is operated by the Association 
of Universities for Research in Astronomy (AURA) under cooperative agreement 
with the National Science Foundation.} 


\begin{abstract} 
Fringing in CCD images is troublesome from the aspect of photometric quality and
image flatness in
the final reduced product. Additionally, defringing during calibration requires 
the inefficient use of time during the night to collect and produce 
a ``supersky" fringe frame. 
The fringe pattern observed in a CCD image for a given near-IR filter is dominated by small 
thickness variations across the detector with a second order effect caused by the wavelength
extent of the emission lines within the bandpass which produce the interference pattern.
We show that essentially any set of emission lines which generally match the wavelength
coverage of the night sky emission lines within a bandpass 
will produce an identical fringe pattern. 
We present an easy, inexpensive, and efficient method which uses a neon lamp as a flat field
source and produces high S/N fringe frames to use for defringing an image during the
calibration process.
\end{abstract}

\section{Introduction}

Fringing in CCD images occurs due to an interference effect similar to Newton's
Rings\footnote{The observed phenomenon of Newton's Rings 
was first described by Robert Hooke in his 1664 book
{\it Micrographia} although it is named for 
Isaac Newton, who was the first to quantitatively analyze the phenomena.}.
The production of constructive and destructive interference patterns can
cause substantial quantum efficiency variations in thinned CCDs as long wavelength
light is multiply reflected between the front and back surfaces. Fringing begins
to be an issue for CCDs when the absorption depth within the silicon 
becomes comparable to the thickness of the CCD. This occurs for optical wavelengths
of $\sim$700 nm or longer for which the light is internally reflected several 
times before finally being absorbed (Lesser 1990; Howell 2006).

During the commissioning of the upgraded Kitt Peak 4-m telescope Mosaic 1.1 imager
in the fall of 2010, we produced neon lamp dome flats capable of
fringe removal during the image reduction process. The 8K x 8K Mosaic 1.1 imager 
(Sawyer et al., 2010) consists of
eight 2048 x 4096 e2v CCD44-82 deep depletion (DD) CCDs placed together in a single 
mosaiced focal plane. The e2v CCDs have low noise and deep pixel
wells ($\ge$ 200,000 electrons) and are subject to fringing in the R, I, and z bands.
Details of the Mosaic 1.1 imager can be found at the Kitt
Peak National Observatory web site\footnote{http://www.noao.edu/kpno/mosaic/mosaic.html}.

To eliminate the need to use valuable observing time during the night to collect a set of deep
sky exposures in order to produce a ``supersky" 
fringe frame, we used afternoon flat field exposures
of a neon lamp to obtain high S/N, fringe frames. The fringe pattern produced in a CCD image
is shown to consist of an average pattern from all the emission lines present within a given
filter bandpass, 
thus the neon emission line source (Ne lamp) provides a match to the fringe pattern
produced by night sky emission on a filter by filter basis. 
The collection of daytime neon fringe flats provides a fast, efficient method to produce fringe
frames and the de-fringed CCD images show a complete removal of the night sky effects.
 
\section{Fringing in Charge-Coupled Devices}

Fringing most notably occurs within a CCD when near monochromatic light is incident on
the detector. Spectroscopic observations are prone to fringing due to their collection of
dispersed light placing individual wavelengths across pixels (e. g., Malumuth et al., 2003).
Direct images also suffer from fringing (e.g., Gullixson 1992), especially for narrow band 
observations, primarily due to the
bright night-sky emission lines from atmospheric OH molecules. Auroral [OI] emission 
near 630 nm can also be strong at times and partake in fringe production.
Reducing fringing in a CCD can be done by controlling the CCD thinning process to a high level
of flatness and by applying an anti-reflection coating to the back of a thinned CCD. 
Both of these solutions 
provide mitigation by reducing the fringe amplitude from near 50\% or more
to 2\% or less. 

Correcting for fringes in near-IR CCD images during the reduction process is, in principle, 
not hard but often takes valuable observing time to obtain a good
fringe frame to use for calibration. Dome flats or twilight flat fields will not contain 
fringes as they are illuminated by polychromatic light from the flat-field lamps or 
the setting (rising) sun. Observers typically depend on observatory 
archive fringe frames or produce fringe
images themselves using a large number of deep exposures taken during the night (e.g., Tyson
1990).
These dithered exposures are then medianed together to remove the stars and combined to
provide a supersky flat (hopefully) containing the fringe pattern of the CCD that can then
be used for fringe removal (Gullixson 1992).

Producing a night time fringe frame is a time consuming procedure both at the telescope
(taking valuable observing time) and in the reduction process (combining the images) and often
only produces a low quality, low S/N  fringe calibration image. 
In addition, the OH emission lines responsible for the near-IR fringing are particularly
troublesome as their intensity can change randomly and rapidly throughout a night and 
can even show periods of
zero intensity, perhaps during the time period when the sky flats are being obtained.

Figure 1 shows the night sky spectrum from 650 to 930 nm (Massey and Foltz 2000) 
and illustrates the
large contribution to the sky flux from OH emission bands. For a given filter such as I or z, 
numerous, essentially monochromatic emission lines combine to cause the CCD fringing. 

The shape and spacing of a CCD fringe pattern is dominated by thickness variations of the 
CCD substrate itself. The variations are not uniform and lead to a fringe pattern 
which is only semi-regular in spacing, shape, and interference band width. 
The wavelengths of the emission lines present within a bandpass account for 
the width of the alternating fringe pattern which would be nicely regular is the CCD thickness
variation was extremely uniform. 
The thickness variation needed for a single fringe to occur, $\Delta$$t$ is given by (Janesick
2000)
$$\Delta t = \frac{\phi \lambda cos \theta}{4 \pi n}$$
where $\phi$ is the phase difference needed for interference (2$\pi$), $\lambda$ is the
wavelength of light, $\theta$ is the angle of incidence (assumed to be normal), 
and $n$ is the index of refraction of silicon taken to be constant for our purposes
with a value of
3.6\footnote{http://www.pveducation.org/pvcdrom/appendicies/optical-properties-of-silicon}.
Using the I bandpass as an example, the change in CCD thickness between where a fringe 
at 750 nm will occur versus
one at 900 nm is 0.021 microns. For a typical modern thinned CCD (such as the e2v device)
the thickness variation over $\sim$2048 pixels will be about 1 micron.
If this were a uniform thickness change, say from the center to the edge of the CCD,
the observed fringe separation at these two emission line wavelengths 
would be $\sim$40 pixels. 
Given that there are OH emission lines covering the entire I (and R and z) 
bandpass, the fringes 
from each individual line blend together and produce bright and dark bands, each 
roughly 40 pixels wide.
Thus, the OH emission lines produce, for practical purposes, a single broad fringe 
pattern on the image consisting of many blended fringes from the various emission lines, each
of which is offset by a small fraction of a pixel ($\sim$0.05 pixels for $\Delta\lambda$=50
\AA). 
Therefore, the average fringe pattern produced on a CCD image 
by the summation of all the OH emission lines
within a given bandpass can be well approximated by essentially any set of emission 
lines that generally have the same wavelength distribution with a given bandpass.

\section{Neon Lamp Flat Field Images}

Since any set of emission lines covering a bandpass will provide the same fringe pattern
for a given filter and CCD combination, 
we examined the use of a neon lamp as a source to provide a fringe image.
The spectrum of a neon lamp in the near-IR is shown in Fig. 1 where it can be seen to provide 
a similar emission line distribution pattern to that of OH within the filters of concern.

A neon lamp test fixture was attached to the front end of the Kitt Peak 4-m 
Mayall telescope during the commissioning run of the Mosaic 1.1 wide-field imaging camera.
Since Mosaic 1.1 is a a prime focus instrument, the top of
the secondary tube was closed and provided a nice mounting surface for our neon paddle lamp
(Fig 2).
The telescope was then pointed at the flat field screen mounted inside the dome, the neon 
lamp alone illuminated the flat field screen, and 
several trial neon fringe flat exposures were obtained. In our case, 
180 sec exposure times provided high S/N fringe flat images in R, I, and z bandpasses 
with our I band neon flat image shown in Fig. 3. 
The neon lamp images were obtained during the afternoon with the dome closed 
using a process identical to the method
used to obtained dome flats. The neon fringe flats were then used as fringe frames during the
image reduction and calibration process.

\section{Defringing}

Mosaic 1.1 images of a star field were obtained in I and z band on a night when
fringing due to OH was readily apparent. The level of the fringing in the I band image 
amounts to a 2-3\% modulation, increasing to near 5\% in z band for these same CCDs 
A similar amplitude increase was observed in the
fringe level between I and z bands in the Gemini-N GMOS e2v DD CCDs (Hook et al., 2004).
Comparison of the fringes in the neon fringe
flat to those caused by OH emission in the I band star field image, show 
an identical pattern of interference bands modulated by the CCD thickness variations 
(Fig. 4). In addition, we note that both fringe patterns show interference bands with
semi-regular spacing near 40 pixels. This match in fringe pattern 
is to be expected as both sources produce a set of 
emission lines covering the I bandpass. 

Following the usual image reduction process (e.g., Newberry 1991; Gullixson 1992; Howell 2006)
and using the neon fringe flat
field to defringe the star image, the final image shows a  a complete removal of 
the night sky fringe pattern (Fig 4). Examination of an average row plot across the
raw image and the same region of the final image is presented in Fig. 5. We see here the
2-3\% fringing present in the raw I band image due to the OH night sky emission lines. 
After reduction, the fringe pattern has been completely removed (Fig. 5) 
and the image is well flattened. The high S/N neon fringe flat has served its purpose well
and provided an easy calibration method for fringe removal.

\section{Conclusion}
We presented an easy and efficient method to provide a high S/N fringe frame using a neon lamp
to illuminate a flat field screen inside a dome. Observations of this illuminated screen,
taken in a similar manner to normal flat field exposures, can be accomplished during the day
obviating the need to use valuable time during the night to produce a supersky frame.
The fringe pattern produced by neon, or essentially any set of similarly distributed
emission lines within a given bandpass, is
identical to that produced by night sky OH emission and therefore can be used to fringe
correct astronomical images of interest. Images containing fringes and corrected using neon
fringe flats show complete fringe removal as well as good flatness leading to high photometric
quality.

The author wishes to thank Dave Sawyer, Heidi Schweiker, and Jim Hutchinson for help during
the neon flat field taking adventure and Rob Seaman for help in retrieving some of the
commissioning images from the NOAO archives.

\clearpage

\begin{figure}
\includegraphics[angle=-90,scale=0.7,keepaspectratio=true]{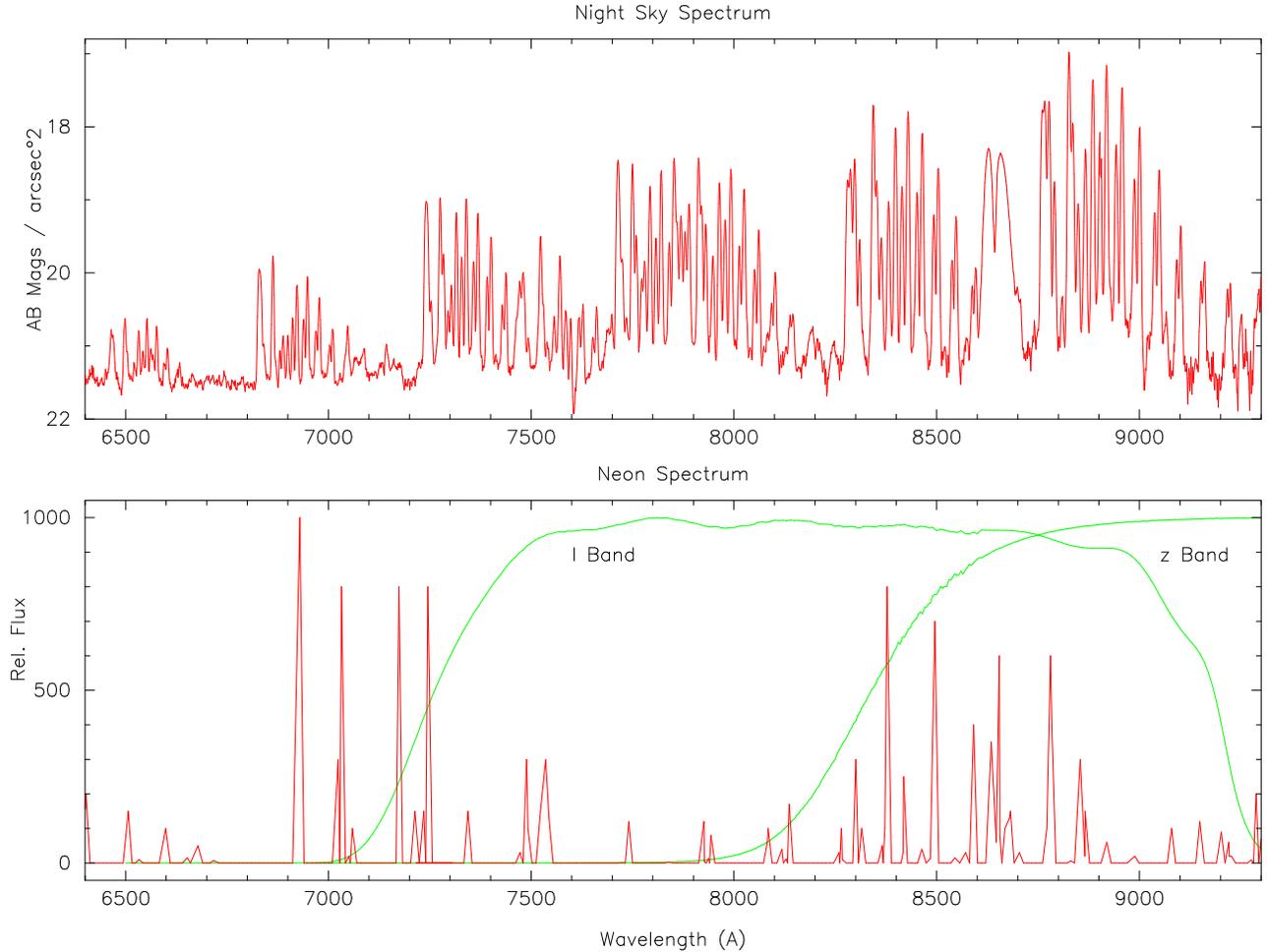}
\caption{Top: Spectrum of the night sky covering the near-IR (Massey \& Foltz 2000). 
Notice the large
emission line bands due to the OH molecule in the Earth's atmosphere. These emission lines 
are the primary cause of CCD fringing.
Bottom: Emission line spectrum of a neon lamp shown over the same spectral region.
Neon also has groups of emission lines within this region which will provide the same fringe
pattern on a CCD as OH emission.
Broadband I and z bandpasses discussed in this paper are indicated.
}
\end{figure}

\begin{figure}
\plotone{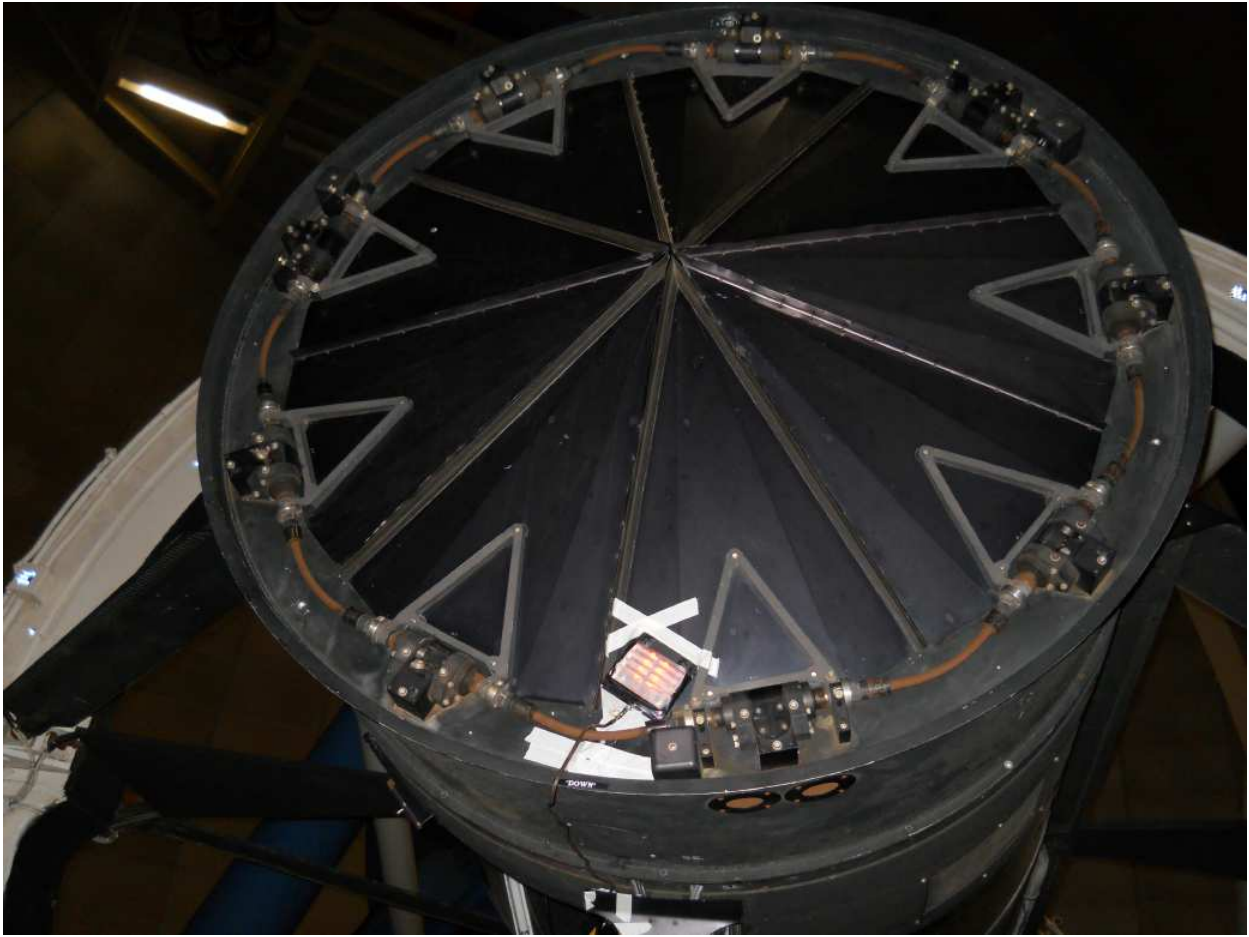}
\caption{Photograph of the neon ``flat field" paddle lamp 
attached to the front end of the secondary
tube on the Kitt Peak 4-m Mayall telescope. This temporary configuration was used to obtain
dome flats with the neon light source illuminating the flat field screen mounted on the 
inside of the 4-m dome.
}
\end{figure}

\begin{figure}
\plotone{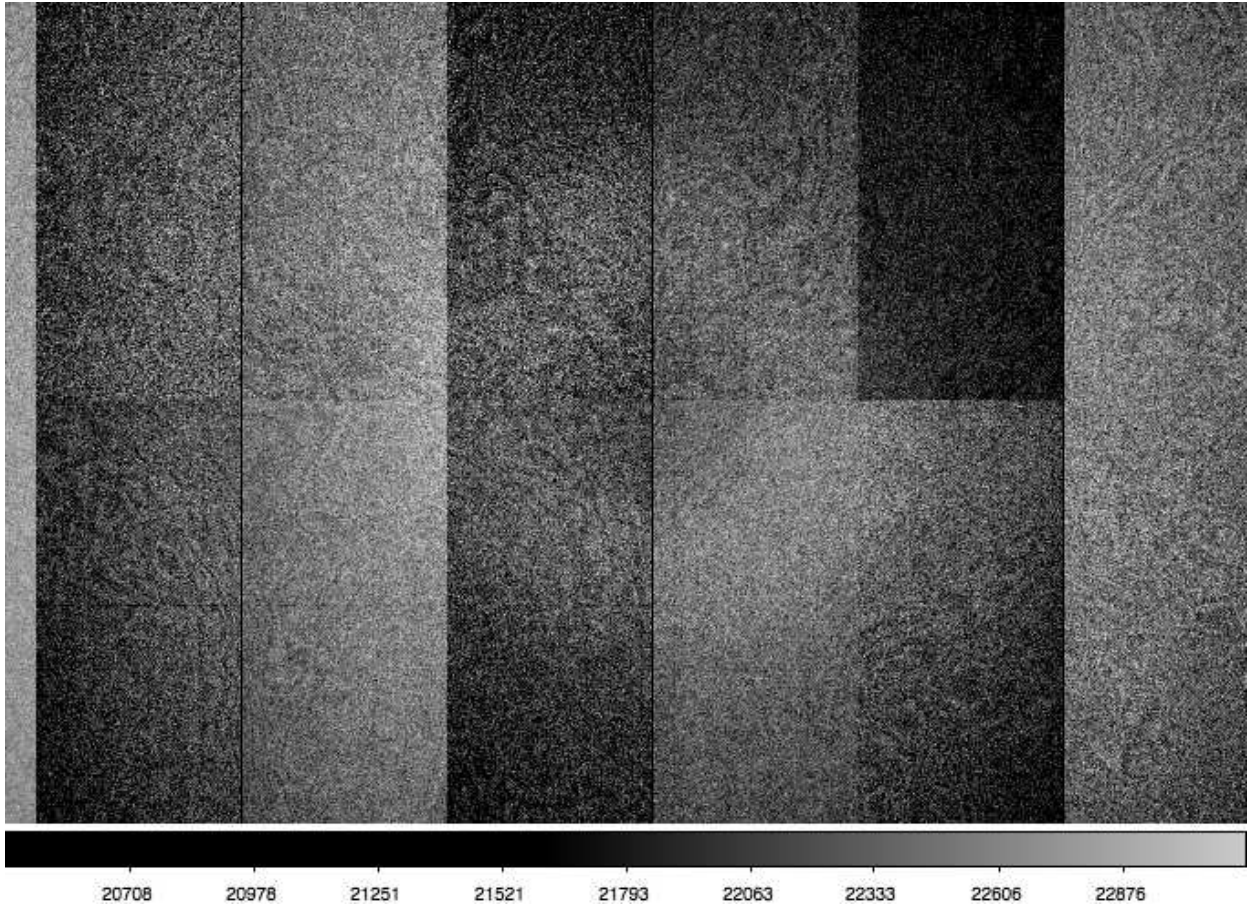}
\caption{The central six CCDs of the Mosaic 1.1 I band neon fringe flat field. This 180 second
exposure shows a raw image where the 12 amplifiers (2 per e2v CCD) are visible due to 
their slightly different bias levels, the 
pupil ghost at the image center, and the strong fringing due to the neon emission
lines in the flat field.
}
\end{figure}

\begin{figure}
\includegraphics[angle=0,scale=0.505,keepaspectratio=true]{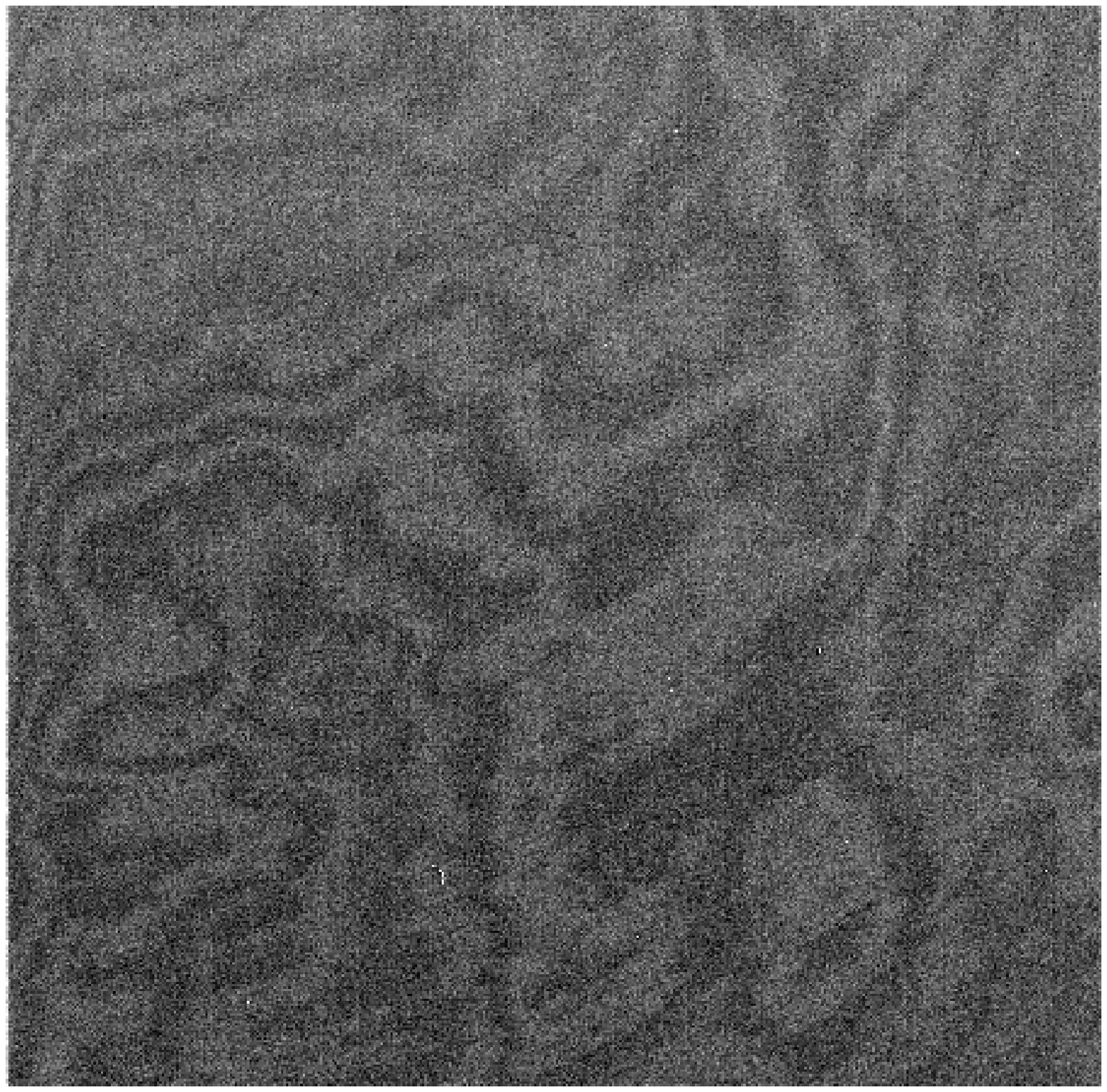}
\includegraphics[angle=0,scale=0.5,keepaspectratio=true]{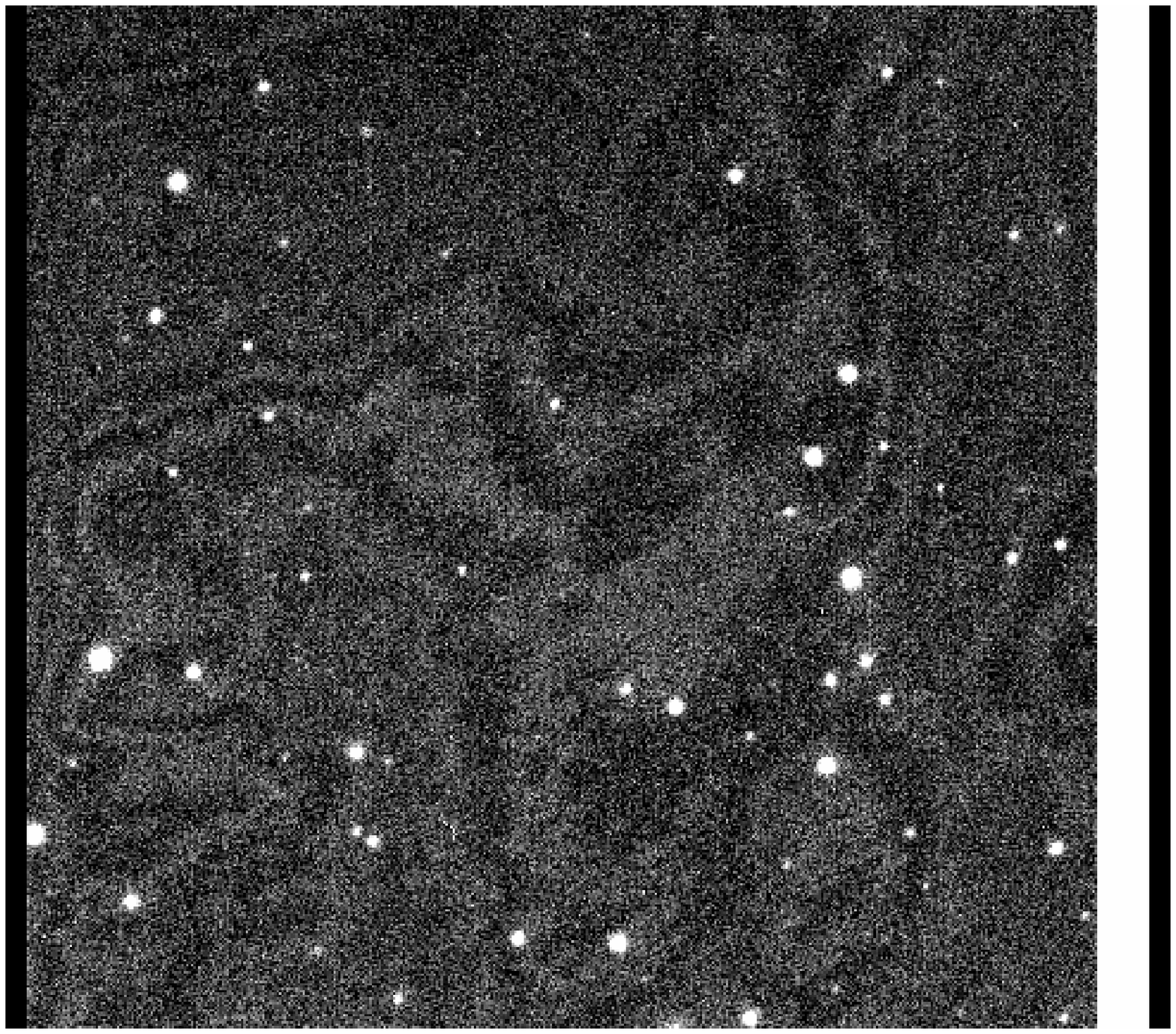}
\includegraphics[angle=0,scale=0.5,keepaspectratio=true]{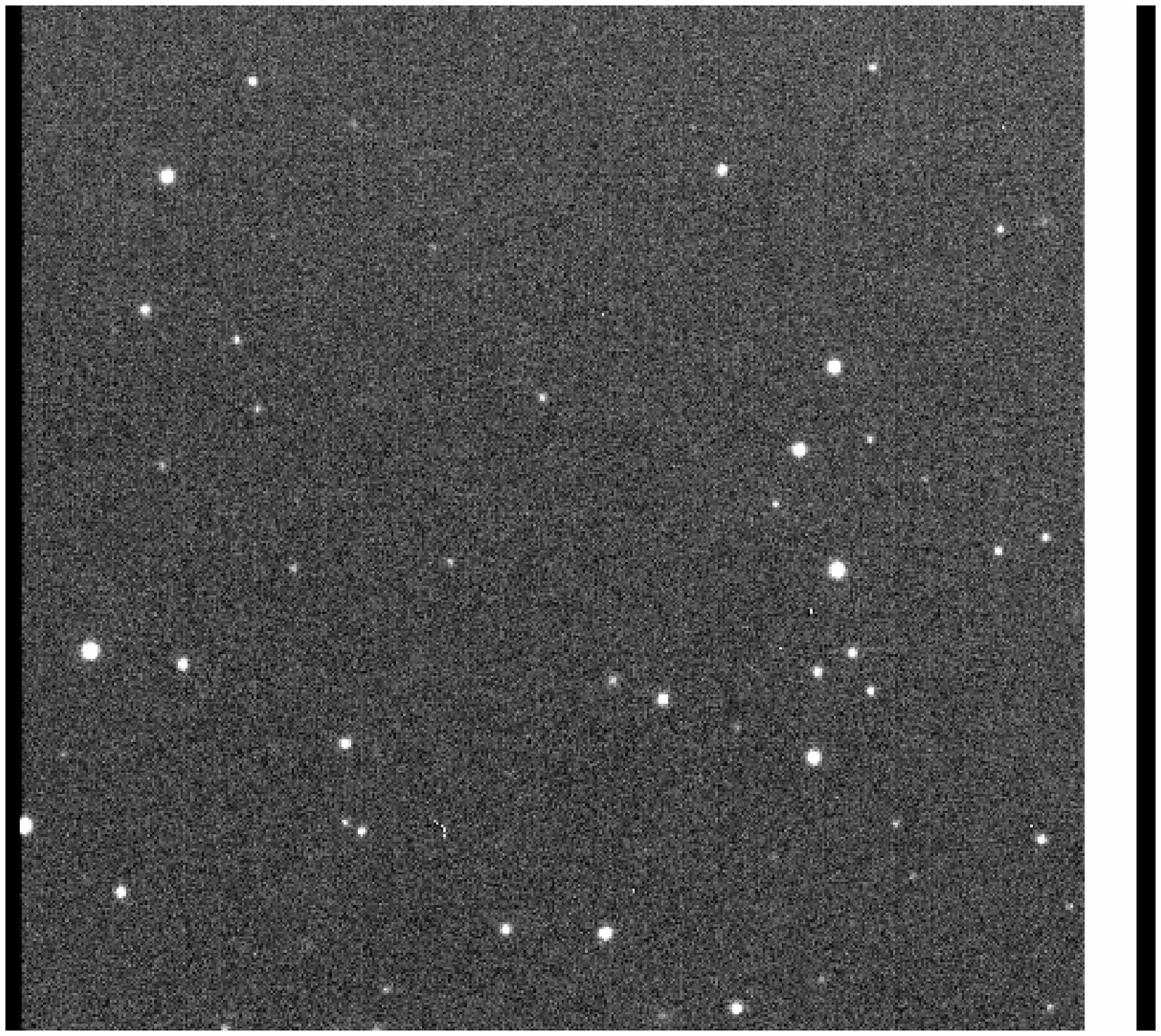}
\caption{A 1024 X 1024  portion of one of the Mosaic 1.1 CCDs showing (TOP)
the I band neon fringe flat field, (MIDDLE) the same CCD section 
showing a night time I band image of a star field taken with strong OH emission, and (BOTTOM)
the same section in the final reduced image showing the complete removal of the CCD fringes.
Note that the fringe pattern is the same in the I band neon fringe flat image and the 
I band night sky image.
}
\end{figure}

\begin{figure}
\includegraphics[angle=0,scale=0.45,keepaspectratio=true]{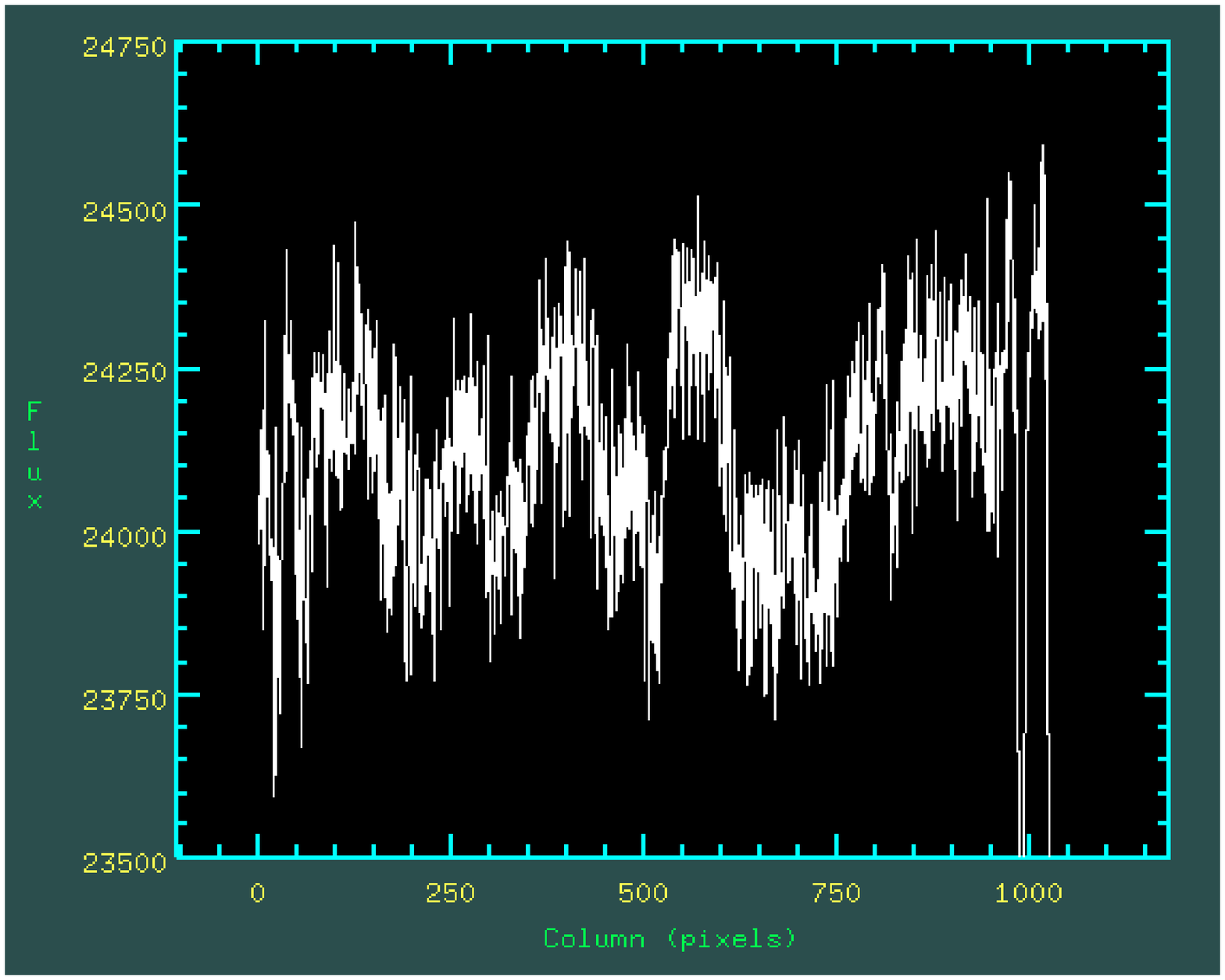}
\includegraphics[angle=0,scale=0.45,keepaspectratio=true]{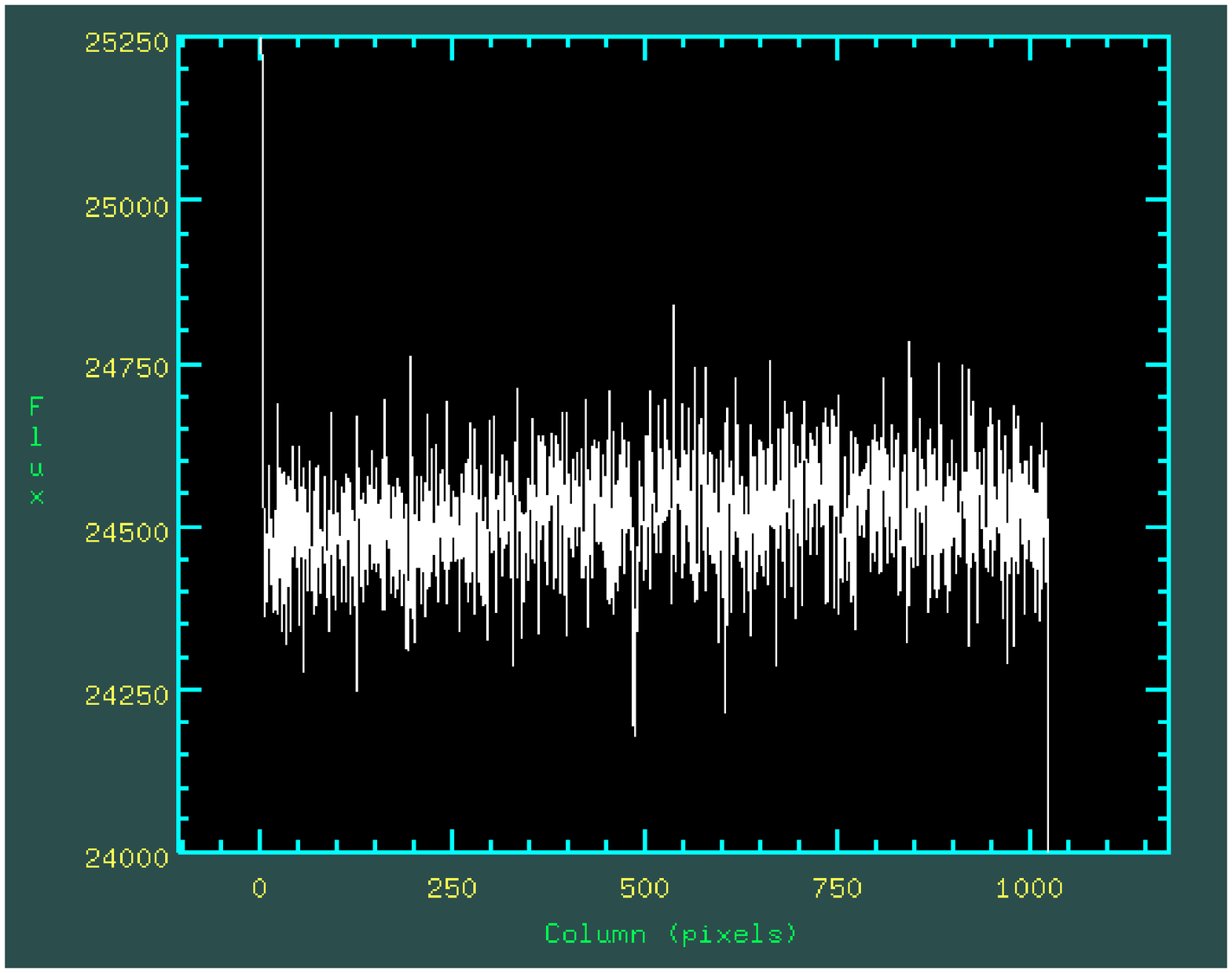}
\caption{Left: A plot of 20 averaged columns across the I band star field image 
shown in Fig. 4.
The fringes in the raw image are obvious and have peak-to-peak amplitudes of 2-3\% (about 500
ADU) and interference spacings near 40-50 pixels.
Right: The same column average plotted after defringing the image using the neon lamp flat 
field. The fringes have been completely compensated for.
}
\end{figure}

\end{document}